Title: Modeling state-transition dynamics in resting-state brain signals by the hidden Markov and Gaussian mixture models

**Authors:** Takahiro Ezaki[1,2], Yu Himeno[3], Takamitsu Watanabe[4,5], and Naoki Masuda[6,7]*

[1]Research Center for Advanced Science and Technology, The University of Tokyo, 4-6-1 Komaba, Meguro-ku, Tokyo 153-8904, Japan

[2]PRESTO, JST, 4-1-8 Honcho, Kawaguchi, Saitama 332-0012, Japan

[3]Department of Aeronautics and Astronautics, The University of Tokyo, 7-3-1 Hongo, Bunkyo-ku, Tokyo 113-8656, Japan

[4]Laboratory for Cognition Circuit Dynamics, RIKEN Centre for Brain Science, Saitama 351-0198, Japan

[5]International Research Center for Neurointelligence, The University of Tokyo 7-3-1 Hongo Bunkyo-ku, Tokyo 113-0033 Japan

[6]Department of Mathematics, State University of New York at Buffalo, Buffalo, NY 14260-2900, USA

[7]Computational and Data-Enabled Science and Engineering Program, State University of New York at Buffalo, Buffalo, NY 14260-5030, USA

* Corresponding author: naokimas@buffalo.edu

**Abstract**

Recent studies have proposed that one can summarize brain activity into dynamics among a relatively small number of hidden states and that such an approach is a promising tool for revealing brain function. Hidden Markov models (HMMs) are a prevalent approach to inferring such neural dynamics among discrete brain states. However, the impact of assuming Markovian structure in neural time series data has not been sufficiently examined. Here, to address this situation and examine the performance of the HMM, we compare the model with the Gaussian mixture model (GMM), which is with no temporal regularization and thus a statistically simpler model than the HMM, by applying both models to synthetic time series generated from empirical resting-state functional magnetic resonance imaging (fMRI) data. We compared the GMM and HMM for various sampling frequencies, lengths of recording per participant, numbers of participants, and numbers of independent component signals. We find that the HMM attains a better accuracy of estimating the hidden state than the GMM in a majority of cases. However, we also find that the accuracy of the GMM is comparable to that of the HMM under the condition that the sampling frequency is reasonably low (e.g., TR = 2.88 or 3.60 s) or the data is relatively short. These results suggest that the GMM can be a viable alternative to the HMM for investigating hidden-state dynamics under this condition.

# INTRODUCTION

Brain dynamics are a product of large-scale networks realized by interaction of functionally specialized regions in the brain (Raichle *et al.*, 2001; Fox *et al.*, 2005; Sporns, 2011; Deco *et al.*, 2015). Such dynamics have been considered to underpin the integration of information (Tononi *et al.*, 1998), cognitive functions (Bressler & Menon, 2010), and their impairments (i.e., neuropsychiatric disorders) (Menon, 2011). Understanding dynamical coordination of brain regions necessitates data-analysis methods that reduce the dimension of large-scale neural data, which are often provided in the form of multivariate time series, without losing much information. Widely used examples include independent component analysis (Calhoun & Adali, 2012) and network analysis (Sporns, 2011; Bassett & Sporns, 2017).

One approach to investigating integrated dynamics of multivariate time-varying neural signals is to assume a relatively small number of latent states and summarize the multidimensional brain activity data at each time point into one of these states. One can estimate time series of the latent state by, for example, the hidden Markov models (HMMs) (Baker *et al.*, 2014; Ryali *et al.*, 2016; Vidaurre *et al.*, 2016, 2017; Taghia *et al.*, 2017; Warnick *et al.*, 2017; Brookes *et al.*, 2018; Nielsen *et al.* 2018; Vidaurre, Hunt, *et al.*, 2018; Vidaurre, Abeysuriya, *et al.*, 2018; Viddaurre, 2021), dynamic functional connectivity (Allen *et al.*, 2014; Calhoun *et al.*, 2014; Nielsen *et al.*, 2018), and energy landscape analysis (Watanabe, Masuda, *et al.*, 2014; Watanabe, Hirose, *et al.*, 2014; Ezaki *et al.*, 2017). This strategy allows us to continue to work on the same time domain as the original data, and therefore to, e.g., compute transition rates between the latent states and interpret state transition events, rather than to reduce the data to static measures (e.g., functional connectivity) or transform the data to the frequency domain implicitly assuming stationarity of the time series. State-transition dynamics have been reported to be closely related to various functions of the brain, including executive function (Ezaki *et al.*, 2018), decision-making (Taghia *et al.*, 2018), and to psychiatric conditions such as autism (Watanabe & Rees, 2017) and schizophrenia (Kottaram *et al.*, 2019). For example, Ezaki et al. (Ezaki *et al.*, 2018) analyzed resting-state functional magnetic resonance imaging (fMRI) data obtained from healthy humans using the energy landscape analysis. They showed that the ease of state transitions between synchronized activity patterns of specific regions of interest (ROIs) explained age-related changes in executive functions. Taghia et al. (Taghia *et al.*, 2018) applied a Bayesian switching linear dynamical systems model to fMRI data obtained from participants performing a $n$-back working memory task. They found a task-specific hidden state and dynamical switching path of the estimated hidden states. Using the energy landscape analysis, Watanabe et al. (Watanabe & Rees, 2017) showed that high functioning autistic adults had atypically stable brain dynamics with lower transition rates among different brain systems and longer dwelling time, and that such over-stability was predictive of both their symptom severity and unique cognitive skills. For resting-state fMRI data obtained from individuals with schizophrenia, Kottaram et al. (Kottaram *et al.*, 2019) reported aberrant transition dynamics among latent states estimated by an HMM. Relevance of hidden states to task-related data in a supervised setting has also been reported. Vidaurre et al. (Vidaurre, Hunt, *et al.*, 2018) found hidden states representing task-related brain states using magnetoencephalography (MEG) data during a button press task.

Among these methods, the HMMs have been widely used for studying fMRI (Ryali *et al.*, 2016; Vidaurre *et al.*, 2017; Warnick *et al.*, 2017; Vidaurre, Hunt, *et al.*, 2018; Vidaurre, 2021) and MEG (Baker *et al.*, 2014; Vidaurre *et al.*, 2016; Taghia *et al.*, 2017; Brookes *et al.*, 2018; Vidaurre, Hunt, *et al.*, 2018; Vidaurre, Abeysuriya, *et al.*, 2018) data recorded from the human brain. The HMM is a model comprising a set of probability distributions of the observables each of which corresponds to a latent (hidden) state and the transition probabilities between the pairs of latent states. By assumption, the state-transition dynamics of an HMM are Markovian, i.e., with no memory effect longer than a single time step. HMMs have been useful in modeling neural dynamics for the following reasons. First, they can be applied to relatively high-dimensional time series (Baker *et al.*, 2014; Brookes *et al.*, 2018; Vidaurre, Hunt, *et al.*, 2018; Vidaurre, Abeysuriya, *et al.*, 2018). Second, they can detect changes in signals without delay in the form of changes in the latent state, which is not straightforward with dynamic functional connectivity calculated with sliding time windows (Vidaurre, Abeysuriya, *et al.*, 2018; Vidaurre, 2021).

The HMM modeling is based on the implicit assumption that the stochastic rule of state transitions depends on the last state and not on the states in the further past. In fact, the practical impacts of this temporal assumption have not been well studied at least with fMRI data. The potential deviation of the data from the assumed Markovian temporal structure may be detrimental to the HMM modeling.

The same analysis pipeline to infer hidden states of the time series data (Fig. 1) can be realized by a simple static mixture model that does not impose any temporal (HMM-like) regularization. Mixture models have been used in neuroimaging research for detecting activation of brain regions (Everitt & Bullmore, 1999; Hartvig & Jensen, 2000) and clustering them into larger ROIs (Górriz *et al.*, 2009). In Everitt & Bullmore (1999) and Hartvig & Jensen (2000), for example, a mixture model with two probability distributions corresponding to activation and deactivation of each voxel, respectively, was used to classify the state of the voxels. The mixture models have also been used for capturing the state transitions in neuroimaging data. For example, Nielsen *et al*. (2018) used mixture models for clustering the dynamic functional connectivity patterns. Vidaurre (2021) proposed mixture models for the principal component analysis (PCA) (as well as their HMM variants) to capture dynamic functional connectivity. Because mixture models do not assume any temporal structure, they are not influenced by the sampling frequency and therefore may serve as useful baseline models with which to assess the efficiency of fitting HMMs. If state transitions in the given data are considerably influenced by the current state, an HMM is expected to perform better than a mixture model. In contrast, if state transitions do not depend on the current state, mixture models may outperform HMMs because of their relative simplicity. In general, complex models may overfit to the data, and their model estimation algorithms are often computationally costly and may end up converging to local optima.

On these grounds, we compare the HMM and Gaussian mixture model (GMM) in the present study. We used the GMM because, in HMMs applied to neuroimaging data, the Gaussian distribution is widely used as the probability distribution conditioned on the hidden state (Baker *et al.*, 2014; Ryali *et al.*, 2016; Vidaurre *et al.*, 2016, 2017; Taghia *et al.*, 2017; Warnick *et al.*, 2017; Brookes *et al.*, 2018; Vidaurre, Hunt, *et al.*, 2018; Vidaurre, Abeysuriya, *et al.*, 2018). Then, if we ignore state-transition dynamics as described by a hidden Markov process, the distribution of signals as estimated by the fitting of an HMM is a GMM.

We compare HMMs and GMMs on synthetic data sets which we generate from human resting-state fMRI data. Empirical resting-state fMRI data vary in the time resolution, corresponding to the TR (i.e., repetition time), the length of the recording time (i.e., number of image volumes), and the number of participants. Therefore, we examine how these three factors influence the performance of the HMMs and GMMs in estimating the hidden states of fMRI signals over time. We show that, while the HMM outperforms the GMM when the sampling frequency is high (e.g., TR = 0.72 sec), the advantage of the HMM diminishes as the sampling frequency decreases (i.e., larger TR). In addition, when the recording time is short (i.e., a small number of volumes or participants), the GMM more robustly estimates the hidden states than the HMM especially when the dimension of the signal is large. Our results provide quantitative guidance on when HMMs work better than GMMs or vice versa.

**MATERIALS AND METHODS**

An overview of the analysis pipeline is shown in Fig. 1.

*Gaussian mixture model*

Assume that there are $t_{\max}$ observations of $N$-dimensional data, $\mathbf{x}_t$ ($t = 1, \ldots, t_{\max}$) (Fig. 1(a)). Our Gaussian mixture model assumes that each observation, $\mathbf{x}_t$, is generated from one of the two Gaussian distributions (Fig. 1(b)). We set the number of states to two for simplicity. The probability density of the observed data conditioned on the state is given by $\mathcal{N}(\boldsymbol{\mu}_{s_t}, \boldsymbol{\Sigma}_{s_t})$, where $s_t \in \{1,2\}$ is the hidden (i.e., latent) state, $\mathcal{N}$ denotes a $N$-dimensional multivariate Gaussian distribution, and $\boldsymbol{\mu}_{s_t}$ and $\boldsymbol{\Sigma}_{s_t}$ are the mean and covariance matrix of the Gaussian distribution under hidden state $s_t$, respectively. The marginal probability distribution of $\mathbf{x}_t$ is given by

$$P(\mathbf{x}_t) = \sum_{s=1}^{2} \pi_s \mathcal{N}(\mathbf{x}_t | \boldsymbol{\mu}_s, \boldsymbol{\Sigma}_s), \tag{1}$$

where $\pi_s$ is the probability that hidden state $s$ is taken. One estimates $\pi_s$, $\boldsymbol{\mu}_s$, and $\boldsymbol{\Sigma}_s$ ($s=1, 2$) by maximizing the log-likelihood function given by

$$\ln P(\mathbf{x}_1, \mathbf{x}_2, \ldots, \mathbf{x}_{t_{\max}} | \boldsymbol{\pi}, \boldsymbol{\mu}, \boldsymbol{\Sigma}) = \sum_{t=1}^{t_{\max}} \ln \left\{ \sum_{s=1}^{2} \pi_s \mathcal{N}(\mathbf{x}_t | \boldsymbol{\mu}_s, \boldsymbol{\Sigma}_s) \right\}. \tag{2}$$

We used the expectation-maximization (EM) algorithm, which is typically used for maximizing Eq. (2) (Dempster *et al.*, 1977; Lindsay, 1995).

Then, the time course of the hidden state, $\hat{s}_t$ ($t = 1, \ldots, t_{\max}$), given the observations is estimated by

$$\hat{s}_t = \arg\max_s P(s|\mathbf{x}_t) = \arg\max_s \frac{\hat{\pi}_s \mathcal{N}(\mathbf{x}_t | \hat{\boldsymbol{\mu}}_s, \hat{\boldsymbol{\Sigma}}_s)}{\sum_{r=1}^{2} \hat{\pi}_r \mathcal{N}(\mathbf{x}_t | \hat{\boldsymbol{\mu}}_r, \hat{\boldsymbol{\Sigma}}_r)}, \tag{3}$$

where $\hat{\pi}_s, \hat{\boldsymbol{\mu}}_s$, and $\hat{\boldsymbol{\Sigma}}_s$ are the maximum likelihood estimator obtained by the EM algorithm.

We analyze the data using the GMM package in *scikit-learn* (Pedregosa *et al.*, 2011). We used the default setting of *scikit-learn* for determining the initial conditions for the EM algorithm.

*Hidden Markov model*

We consider HMMs with Gaussian components (Ephraim & Merhav, 2002) (Fig. 1(c)). The model assumes that each of the $N$-dimensional observations $\mathbf{x}_t$ ($t = 1, \cdots, t_{\max}$) is generated from one of the two Gaussian distributions, as in GMMs, and that $s_t$ ($t = 1, \cdots, t_{\max}$) obeys first-order Markovian dynamics given by

$$P(s_t | s_1, \cdots, s_{t-1}) = P(s_t | s_{t-1}). \tag{4}$$

To estimate the HMM, we used an EM algorithm known as the Baum-Welch algorithm (Baum *et al.*, 1970). We used the Viterbi algorithm (Viterbi, 1967) to estimate the time course of the hidden state given the observations. We estimated HMMs for our data using a python package *hmmlearn* (https://hmmlearn.readthedocs.io/), which was originally developed as part of *scikit-learn* (Lindsay, 1995). We used the default setting of *hmm-learn* for determining the initial conditions for the EM algorithm.

The EM algorithm does not guarantee the exact optimization due to local minima. Therefore, for both GMM and HMM, we carried out the optimization procedure 10 times and adopted the model that attained the largest likelihood. In the present study, all the estimations successfully converged.

*Resting-state fMRI data*

We used "S1200 extensively processed rfMRI data" provided by the Human Connectome Project (HCP, https://www.humanconnectome.org/) (Van Essen *et al.*, 2012). The data collection was approved by the local ethics committees that participated in HCP, and all participants provided written consent. The data set includes the group independent component analysis (ICA) data for 1003 participants (22–35yo, 534 females). In this section we briefly describe the preprocessing procedure to obtain this data set. See Smith *et al.* (2013), Griffanti *et al.* (2014), and Salimi-Khorshidi *et al.* (2014) for the details of the preprocessing.

The participants completed four sessions of 15-min echo planar imaging (EPI) sequence on a 3T Siemens Connectome-Skyra (TR = 0.72 s, TE = 33.1 ms, 72 slices, 2.0 mm isotropic; FOV = 208 × 180 mm) and a single T1-weighted sequence (TR = 2.4 s, TE = 2.14 ms, 0.7 mm isotropic, FOV = 224 × 224 mm). Each session yielded 1200 volumes (i.e., observations) of EPI images. The fMRI data were first minimally preprocessed according to Smith *et al.* (2013). Then, artifacts were removed using ICA+FIX (Griffanti *et al.*, 2014; Salimi-Khorshidi *et al.*, 2014) and inter-participant registration of cerebral cortex using MSMAll (Robinson *et al.*, 2014; Glasser *et al.*, 2016). Then, the group-ICA was performed by the MELODIC's incremental group PCA algorithm (Hyvärinen, 1999; Beckmann & Smith, 2004). The group-ICA was carried out for dimensions (i.e., number of independent components) $N$ = 15, 25, 50, 100, 200, and 300. We did not use group-ICA data generated with $N$ =100, 200, and 300 because of high computational cost for estimating the models used in the present study.

We fed the fMRI data to algorithms for estimating GMMs or HMMs after concatenating the observed signals obtained from all the sessions from a single participant into one sequence and then the sequences obtained from all the participants into one sequence. In the concatenated data, the final volume of sessions 1, 2, and 3 is followed by the first volume of sessions 2, 3, and 4, respectively, although different sessions are not causally related to each other. In practice, the influence of the concatenation on the estimation of the HMM is considered to be negligible because each session is sufficiently long (i.e., 1200 volumes) (Vidaurre *et al.*, 2017; Vidaurre, Abeysuriya, *et al.*, 2018).

*GMM-based synthetic time series*

We prepared synthetic fMRI data with an underlying hidden-state dynamics for each of $N$ (= 15, 25 and 50) as follows. First, we fit the GMM with 2 hidden states to the entire $N$-dimensional fMRI data that are composed of 1200 volumes × 4 sessions × 1003 participants. As a result, we obtained the mean vector and covariance matrix of each Gaussian distribution, which we denote by $\hat{\boldsymbol{\mu}}_k^{\text{GMM}}$ and $\hat{\boldsymbol{\Sigma}}_k^{\text{GMM}}$ ($k = 1,2$), and the probability of appearance of each state, $\pi_k$ (see Eq. (1)). Then, by maximizing the likelihood given by Eq. (2) given the estimated model, we estimated the hidden states for the first session (but not for the second to fourth sessions) of the fMRI recording, which we denote by $\hat{s}_{p,t}^{\text{GMM}} \in \{1,2\}$ (with $t = 1, \ldots, 1200$), for each participant $p$ (with $p = 1, \ldots, 1003$).

Second, using $\hat{s}_{p,t}^{\text{GMM}}$, $\hat{\boldsymbol{\mu}}_k^{\text{GMM}}$, and $\hat{\boldsymbol{\Sigma}}_k^{\text{GMM}}$ ($k = 1, 2$), we generated $N$ (= 15, 25, or 50) dimensional synthetic signals, $\mathbf{x}_{p,t}^{\text{GMM}}$ for $p = 1, \ldots, 1003$ and $t = 1, \ldots, 1200$, i.e.,

$$\mathbf{x}_{p,t}^{\text{GMM}} \sim \mathcal{N}(\hat{\boldsymbol{\mu}}_k^{\text{GMM}}, \hat{\boldsymbol{\Sigma}}_k^{\text{GMM}}), \qquad k = \hat{s}_{p,t}^{\text{GMM}}, \tag{5}$$

where $\mathcal{N}(\boldsymbol{\mu}, \boldsymbol{\Sigma})$ denotes the Gaussian distribution with mean vector $\boldsymbol{\mu}$ and covariance matrix $\boldsymbol{\Sigma}$.

Then, we subsampled the signals generated by Eq. (5) to prepare synthetic signals with typical lengths of a resting-state fMRI recording, i.e., $T$ = 5, 10, and 14.4 min, $n_{\text{pat}}$ participants, and a sampling frequency of TR = 0.72, 1.44 2.16, 2.88, and 3.60. Note that $T$ = 14.4 min is the original length of a single session in the HCP data. We used the subsampled data to test the performance of the GMM and HMM.

We generated synthetic fMRI data with the given TR value and the number of participants, $n_{\text{pat}}$, by subsampling as follows. Consider the case $T$ = 10 min. First, for each participant $p$ (with $p = 1, \ldots, n_{\text{pat}}$), the samples $\mathbf{x}_{p,t}^{\text{GMM}}$ ($t = 1, 2, \ldots, 833$) with the associated hidden state labels $\hat{s}_{p,t}^{\text{GMM}}$ provide test data with TR = 0.72 s. Note that, given TR = 0.72 s, $t = 833$ corresponds to 10 min starting from the first volume, i.e., $t = 1$. To generate test data with TR = 1.44 s, we use subsampled data $\mathbf{x}_{p,1}^{\text{GMM}}, \mathbf{x}_{p,3}^{\text{GMM}}, \mathbf{x}_{p,5}^{\text{GMM}}, \ldots, \mathbf{x}_{p,833}^{\text{GMM}}$, which is of length 416. Similarly, the subsampled series $\mathbf{x}_{p,1}^{\text{GMM}}, \mathbf{x}_{p,4}^{\text{GMM}}, \ldots, \mathbf{x}_{p,832}^{\text{GMM}}$ defines synthetic data of length 278 with TR = 2.16 s; the series $\mathbf{x}_{p,1}^{\text{GMM}}, \mathbf{x}_{p,5}^{\text{GMM}}, \ldots, \mathbf{x}_{p,833}^{\text{GMM}}$ defines synthetic data of length 209 with TR = 2.88 s; the series $\mathbf{x}_{p,1}^{\text{GMM}}, \mathbf{x}_{p,6}^{\text{GMM}}, \ldots, \mathbf{x}_{p,831}^{\text{GMM}}$ defines synthetic data of length 167 with TR = 3.60 s. We apply the same subsampling method to generate synthetic data with a range of TR when the assumed recording time is different (i.e., $T$ = 5 or 14.4 min). For example, if $T$ = 5 min, which corresponds to $t = 416$, the data for TR = 1.44 s are given by $\mathbf{x}_{p,1}^{\text{GMM}}, \mathbf{x}_{p,3}^{\text{GMM}}, \ldots, \mathbf{x}_{p,416}^{\text{GMM}}$.

*HMM-based synthetic time series*

We also used another set of synthetic time series data, which we call the HMM-based synthetic time series. Similarly to the generation of the GMM-based synthetic time series, we fitted the HMM with 2 hidden states and the corresponding Gaussian distributions to the entire fMRI data composed of 1200 volumes × 4 sessions × 1003 participants. The fitting yielded the mean vector and covariance matrix of the $N$-dimensional Gaussian distribution corresponding to each state, which we denoted by $\hat{\boldsymbol{\mu}}_k^{\text{HMM}}$ and $\hat{\boldsymbol{\Sigma}}_k^{\text{HMM}}$ ($k = 1, 2$), the transition rates between the two hidden states, and the probability distribution of initial state. Using the estimated HMM and Viterbi algorithm (Viterbi, 1967), we estimated the hidden state labels for the first session of each participant, which we denote by $\hat{s}_{p,t}^{\text{HMM}}$ (with $p = 1, \ldots, 1003$ and $t = 1, \ldots, 1200$). The rest of the procedure is same as that for the GMM-based synthetic time series.

## RESULTS

### *Hidden states estimated from the fMRI data*

First, we compared between the GMM and HMM the Gaussian distributions and the time courses of the hidden state inferred from the empirical fMRI data. The Gaussian distributions estimated for the GMM, parametrized by $\hat{\boldsymbol{\mu}}_k^{\text{GMM}}$ and $\hat{\boldsymbol{\Sigma}}_k^{\text{GMM}}$ (with $k = 1$, 2), and those for the HMM, parametrized by $\hat{\boldsymbol{\mu}}_k^{\text{HMM}}$ and $\hat{\boldsymbol{\Sigma}}_k^{\text{HMM}}$, were similar to each other, but the means estimated by the HMM were farther from 0 than those estimated by the GMM (Fig. 2(a)). Figure 2(b) compares the estimated time courses of the hidden state for a single participant between the GMM and HMM. Although the hidden state is the same between the GMM and HMM in a majority of times (77% for $N = 15$; 77% for $N = 25$; 82% for $N = 50$), the GMM yielded more frequent state flips (the fraction of times at which the state flipped was 0.17 for $N = 15$, 0.19 for $N = 25$, and 0.17 for $N = 50$) than the HMM (0.05 for $N = 15$; 0.04 for $N = 25$; 0.04 for $N = 50$). Consistent with this observation, the probability distribution of the duration of each hidden state has a higher mass at small values for GMMs than HMMs (Fig. 2(c)). Figure 2(c) also indicates that the distribution of the duration of each hidden state has a peak for the HMM but not for the GMM.

In the next subsections, we compare the performance of the GMM and HMM using the two types of synthetic data (i.e., the GMM-based synthetic time series and the HMM-based synthetic time series).

### *Results for the GMM-based synthetic time series.*

We fit the GMM and HMM to GMM-based synthetic time series with various TR values, three values of the recording time per participant (i.e., $T$), various numbers of participants pooled (i.e., $n_{\text{pat}}$), and three values of $N$, and compared the accuracy of fitting. Note that the true hidden state at each time is available in this numerical experiment. We defined the accuracy of estimation as the fraction of times at which the estimated state is the same as the true state. We show the accuracy of estimation for $T = 10$ min in Fig. 3. For both GMM and HMM, the accuracy was low for a small number of participants ($n_{\text{pat}} < 10$), increased as $n_{\text{pat}}$ increased, and saturated before $n_{\text{pat}} \approx 20$ at $> 83\%$. In general, more participants were required for attaining a given accuracy value when $N$ is larger or the sampling frequency is lower (i.e., the TR is larger). For the HMM but not for the GMM, the accuracy increased as the sampling frequency increased. When the sampling frequency is high (i.e., small TR) and $n_{\text{pat}}$ is relatively large, the HMM outperformed the GMM (Fig. 3(a)). However, this advantage of the HMM diminishes when the TR is large, $n_{\text{pat}}$ is small, or $N$ is large (Fig. 3(a)–(e)). For example, when $N = 25$ and TR = 2.88 s, the GMM outperformed the HMM for $1 \le n_{\text{pat}} \le 11$ (Fig. 3(e)).

For $T = 5$ and 14.4 min, the results were qualitatively the same as those for $T = 10$ min (Supporting Information, Figs. S1 and S2). Because $T$ is proportional to the length of the data as $n_{\text{pat}}$ is, the accuracy is higher for a larger $T$ value. It should also be noted that, with $T = 5$ min, the GMM outperformed the HMM in wider parameter regions than with $T = 10$ min.

### *Results for the HMM-based synthetic time series*

We applied the same analysis as that in the previous section to HMM-based synthetic time series. The accuracy of estimating the true hidden state is shown in Fig. 4 for the GMM and HMM estimators, various values of TR, $T = 10$ min, various values of $n_{\text{pat}}$, and three values of $N$. The results were similar to those obtained for the GMM-based synthetic time series. When TR is large, $N$ is large, and $n_{\text{pat}}$ is small, the accuracy of estimation for the GMM was comparable to that for the HMM. Otherwise, the HMM outperformed the GMM. With the HMM-based synthetic time series, the HMM yielded a higher accuracy than the GMM in a wider parameter region than with the GMM-based synthetic time series.

We also carried out the same analysis with $T = 5$ and 14.4 min to confirm that the results were qualitatively the same as those with $T =$ 10 min and that the accuracy was larger for a larger $T$ in general (Supporting Information, Figs. S3 and S4).

# DISCUSSION

We compared the performance of estimating the time course of the hidden state between the GMM and HMM. On the two types of synthetic time series that we tested, the HMM inferred the true hidden state more accurately than the GMM in the majority of cases. However, when the length (i.e., $n_{\text{pat}} \times T$) and sampling frequency of the data are limited, the accuracy of estimation for the GMM was comparable to or even better than that for the HMM. The performance of the GMM and HMM estimators also depended on the underlying dynamics of the hidden state, i.e., the type of synthetic time series.

The accuracy of estimating the hidden state using HMMs depended on the sampling frequency of the data. In practice, HMMs are efficient with a reasonable sampling frequency with which the contribution of the first-order dynamics dominates (Martinez-Diaz *et al.*, 2007). In our results, the HMM outperformed the GMM in the case of the highest sampling frequency (i.e., TR = 0.72 s) on both the GMM- and HMM-based synthetic time series. Moreover, for the HMM-based synthetic time series, the accuracy score for the HMM was larger when TR was smaller (blue lines in Fig. 3). These results suggest that the HMM will probably perform better than the GMM if TR is less than 0.72 s. Note that TR = 0.72 s or similar sampling frequencies are becoming common in fMRI experiments. Furthermore, with TR = 2.88 and 3.60 s, which also fall in a range of TR that remains common in human fMRI experiments, the HMM and GMM performed similarly given a sufficient amount of data. This result suggests that, in this situation, we may benefit from using the GMM, which is conceptually, mathematically, and algorithmically simpler than the HMM.

When the given data are short and sampled with a low frequency (e.g., TR = 3 s), GMMs are probably more advantageous than HMMs in that a GMM with the same number of components as an HMM is better at avoiding overfitting. Furthermore, GMMs seem to be also advantageous over HMMs when the given data are long or the estimation procedure has to be run many times. Up to our numerical effort, estimation of GMMs was at least a couple of times faster than that of HMMs including when we attempted to estimate models with more than two states. This observation is consistent with a previous study reporting that estimation of the HMM is often computationally challenging when the data is long or the number of participants is large (Vidaurre, Abeysuriya, *et al.*, 2018).

We refrained from optimizing the number of hidden states. This is because there is no established way to do so (Celeux & Durand, 2008; Pohle *et al.*, 2017), although some methods based on, e.g., the free energy in the VB algorithms (Rezek & Roberts, 2005) and the number of appearances of each state in the estimated hidden-state time courses (Vidaurre, Hunt, *et al.*, 2018) have practically been used. (Also see Nielsen *et al.* (2018) for an infinite HMM approach, which finds the appropriate number of hidden states but is not computationally feasible in many practical cases.) In previous studies using fMRI data, the estimated numbers of hidden states are distributed in a wide range, i.e., between 5 and 19 (Ryali *et al.*, 2016; Vidaurre *et al.*, 2017; Vidaurre, Abeysuriya, *et al.*, 2018; Kottaram *et al.*, 2019; Scofield *et al.*, 2019; Stevner *et al.*, 2019). In contrast to these studies, we assumed two hidden states for the sake of simplicity. This choice was also motivated by a previous study reporting that the hidden states were robustly agglomerated into two clusters in human fMRI data and that the frequency of the two states was heritable and related to cognitive measures (Vidaurre, Hunt, *et al.*, 2018). We also found in our previous work with energy landscape analysis that transitions among two or three macroscopic states were correlated with participants' behavior in a bistable visual perception task (Watanabe, Masuda, *et al.*, 2014) and executive function (Ezaki *et al.*, 2018). Therefore, we believe that characterizing brain dynamics by transitions among an *a priori* determined small number of states, as we have done in the present study, is a useful approach.

The performance of the GMMs and HMMs may depend on various factors which we have not examined in this study, such as the preprocessing method, scanner, and tasks. In particular, the type of task is expected to modulate the hidden-state dynamics and may inform the choice of the model with state dynamics. We observed that the duration of the hidden state is qualitatively different between the GMM and HMM (Fig. 3(c)). Note that various generalizations of the HMM (i.e., hidden semi-Markov models) (Yu, 2010) have been proposed to incorporate different types of the distribution of the duration. In the absence of the ground-truth data for the hidden state dynamics, which is generally the case, we do not have a particular hypothesis regarding the shape of the distribution of the duration of the hidden state. Instead, in this study we focused on the effect of other key parameters on the estimation accuracy of the GMM and HMM. Additional information that accompanies recorded brain signals, such as behavioral switches during a task, may help us to better compare the GMM, HMM, and their variants.

**ACKNOWLEDGEMENTS**

TE acknowledges the support provided through JST PRESTO (No. JPMJPR16D2) and JSPS Grant-in-Aid for Scientific Research (No. 20H01789). Data were provided by the Human Connectome Project, WU-Minn Consortium (Principal Investigators: David Van Essen and Kamil Ugurbil; 1U54MH091657) funded by the 16 NIH Institutes and Centers that support the NIH Blueprint for Neuroscience Research; and by the McDonnell Center for Systems Neuroscience at Washington University.

**CONFLICT OF INTEREST**

Authors declare no conflict of interest.

**AUTHOR CONTRIBUTIONS**

TE and NM designed the research. TE and YH performed numerical experiments. TE, TW and NM discussed the results. TE, TW and NM wrote the paper.

**DATA AVAILABILITY STATEMENT**

The data used in this study are publicly available (Human Connectome Project: https://www.humanconnectome.org/).

**ABBREVIATIONS**

EM = expectation maximization

EPI = echo planar imaging

fMRI = functional magnetic resonance imaging

GMM = Gaussian mixture model

HCP = Human Connectome Project

HMM = hidden Markov model

ICA = independent component analysis

MEG = Magnetoencephalography

ROI = region of interest

TR = repetition time

PCA = principal component analysis

**FIGURES**

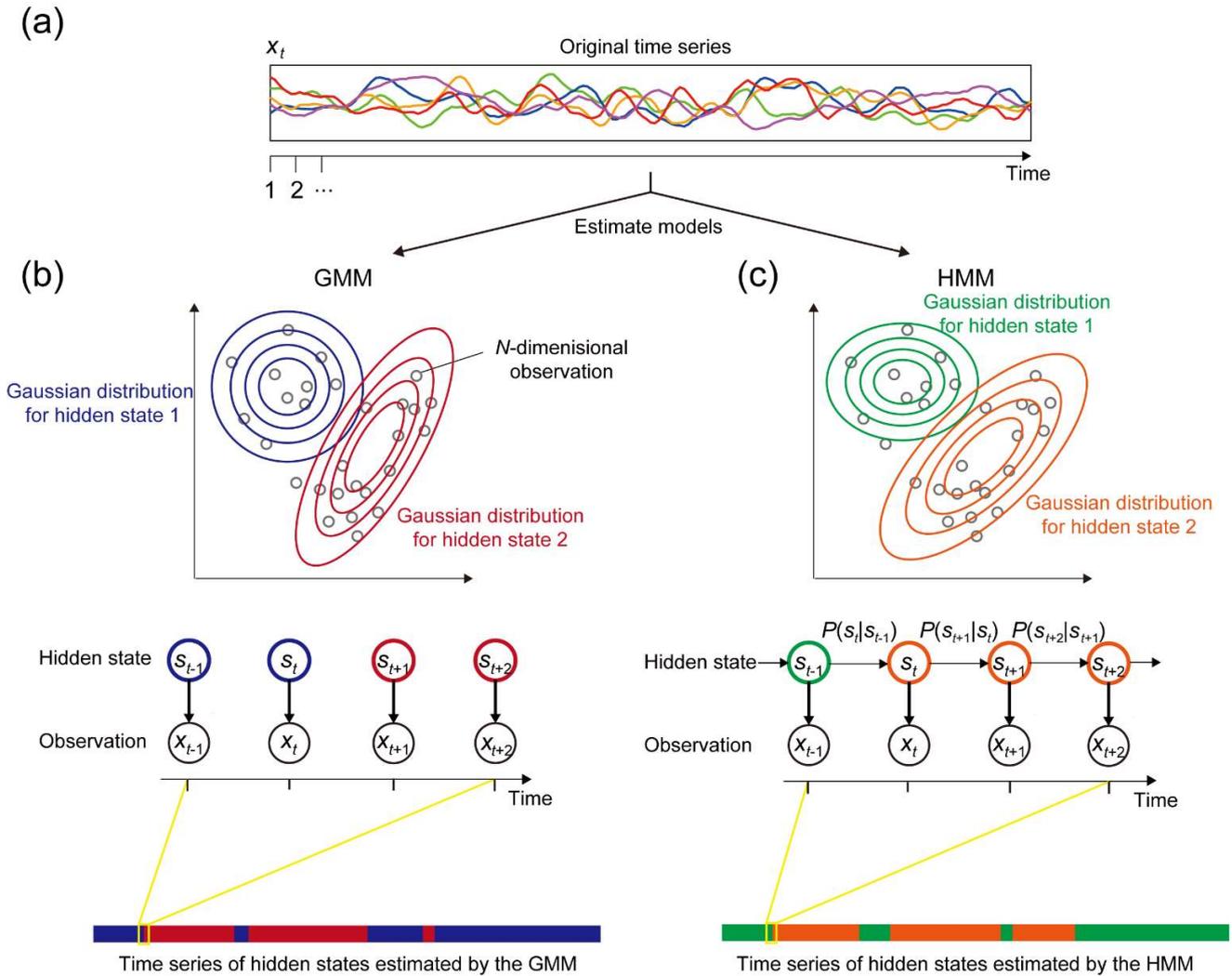

**Figure 1: Overview of the estimation of hidden-state dynamics using GMMs and HMMs.** (a) A multivariate time series in discrete time such as fMRI data. (b) One fits a GMM with two components to the multivariate time series data shown in (a). The case of $N = 2$ is schematically shown. The estimation of the GMM enables us to associate one of the hidden states (shown in color) to the data point at each discrete time, $\mathbf{x}_t$. Using the estimated GMM, one can estimate the time course of the hidden state. (c) One fits an HMM with two hidden states to the same data. In general, how the data points are clustered into two hidden states is different between the GMM and HMM. Using the estimated HMM, one can estimate the time course of the hidden state.

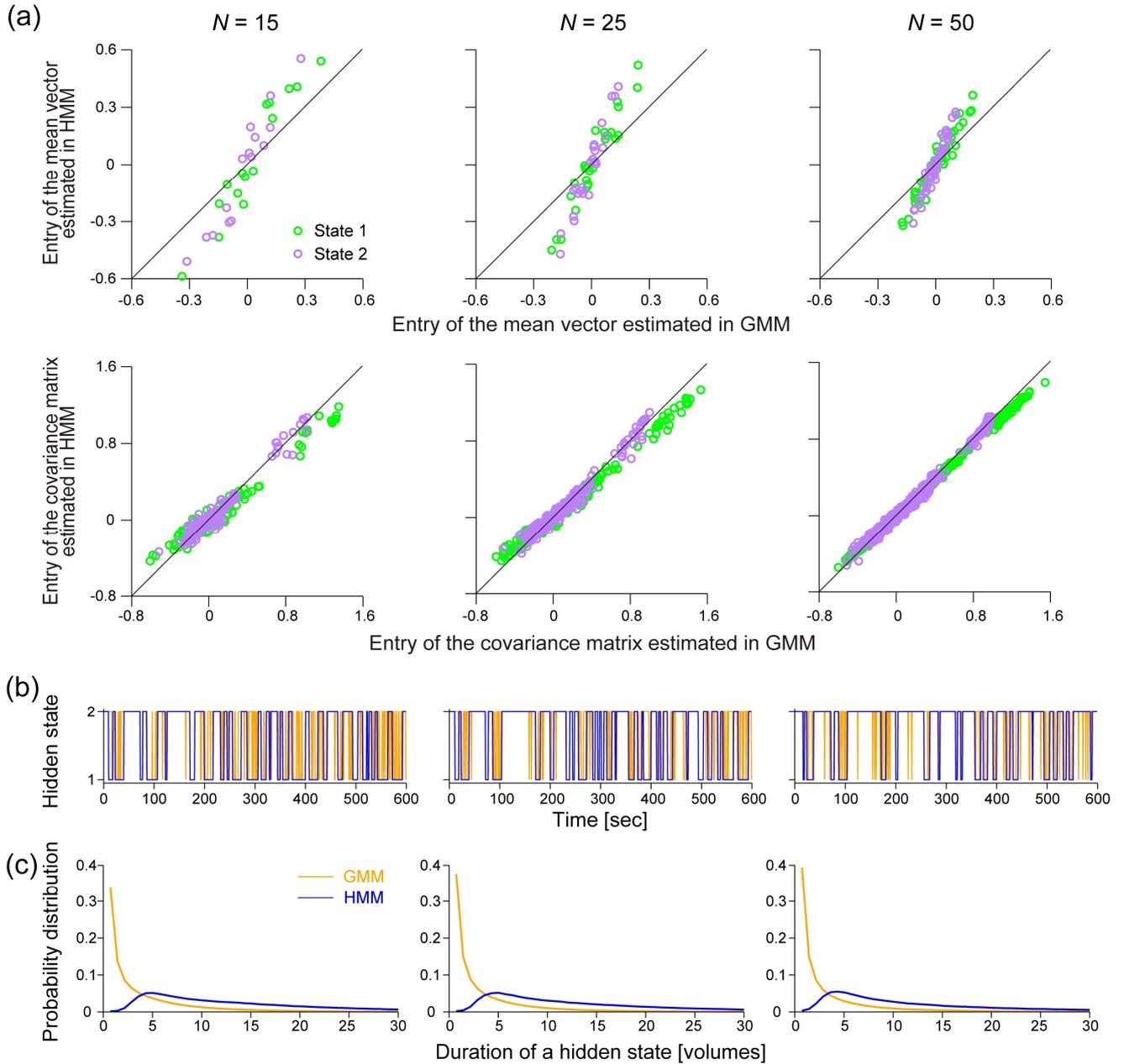

**Figure 2: Comparisons between the GMM and HMM fitted to the fMRI data.** (a) Mean vectors ($\hat{\boldsymbol{\mu}}_k^{\text{GMM}}$ vs $\hat{\boldsymbol{\mu}}_k^{\text{HMM}}$) and covariance matrices ($\hat{\boldsymbol{\Sigma}}_k^{\text{GMM}}$ vs $\hat{\boldsymbol{\Sigma}}_k^{\text{HMM}}$). A circle represents each entry of the mean vector or covariance matrix in the estimated GMM and HMM. The solid lines represent the diagonal. (b) A sample time course of the estimated hidden state labels, $\hat{s}_{p,t}^{\text{GMM}}$ and $\hat{s}_{p,t}^{\text{HMM}}$, for 10 min and a single participant. (c) Distribution of the duration of a hidden state, computed based on the entire sequences of the hidden state, $\hat{s}_{p,t}^{\text{GMM}}$ and $\hat{s}_{p,t}^{\text{HMM}}$ ($1 \leq p \leq 1003, 1 \leq t \leq 1200$).

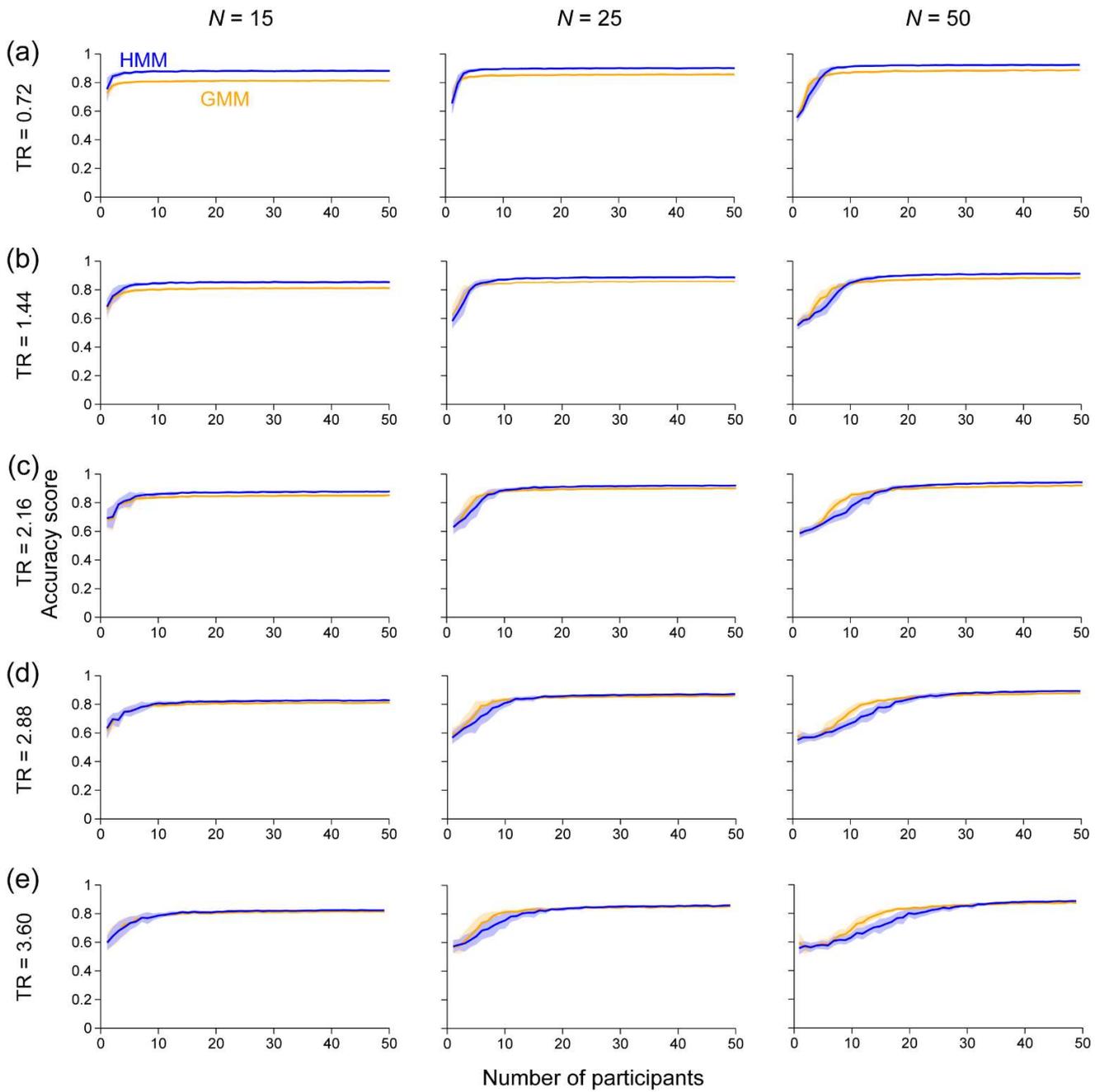

**FIGURE 3: Accuracy of estimating the hidden state for the GMM-based synthetic time series for *T* = 10 min.** (a) TR = 0.72. (b) TR = 1.44. (c) TR = 2.16. (d) TR = 2.88. (e) TR = 3.60. The shaded regions represent one standard deviation.

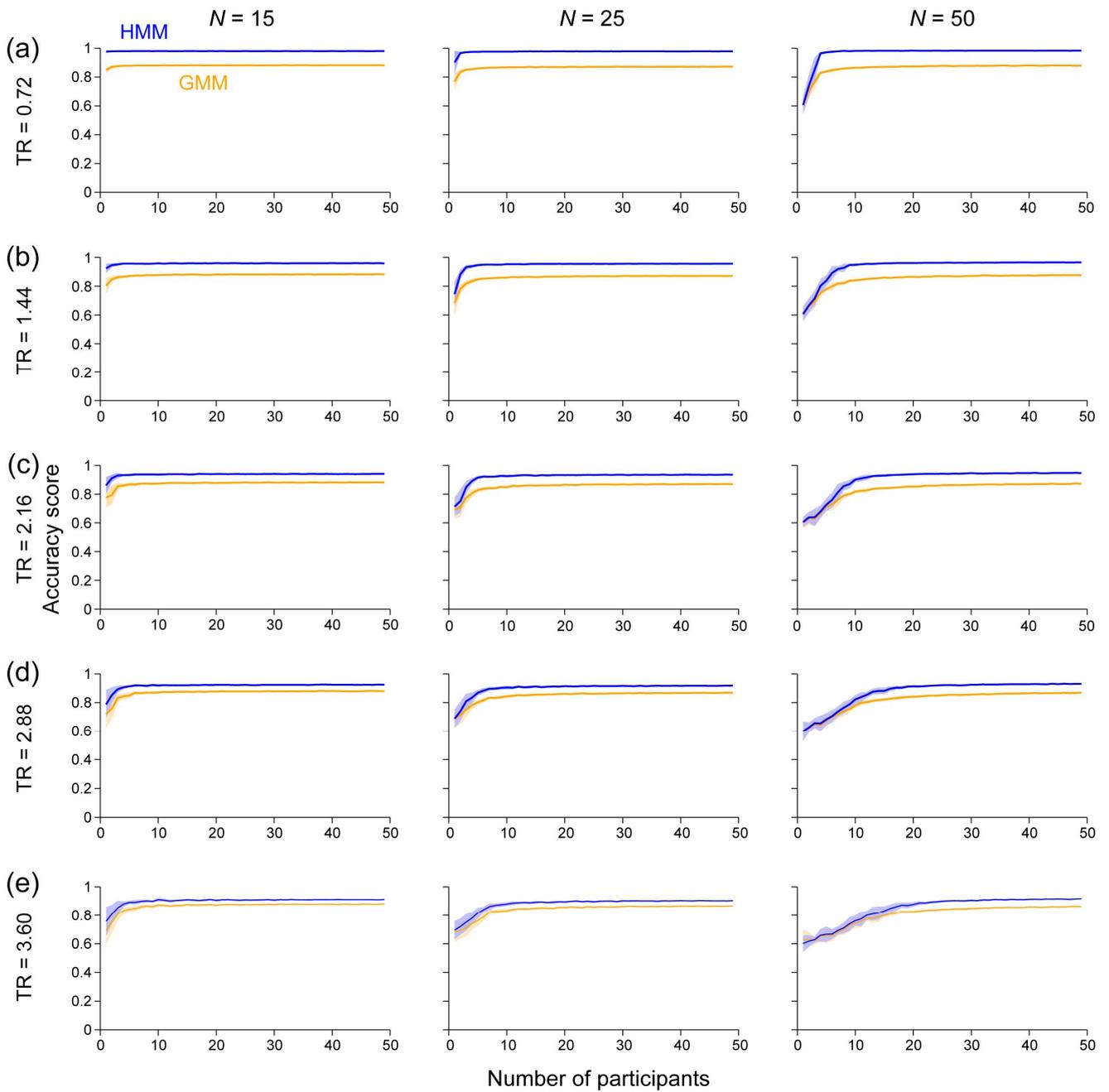

**FIGURE 4: Accuracy of estimating the hidden state for the HMM-based synthetic time series for *T* = 10 min.** (a) TR = 0.72. (b) TR = 1.44. (c) TR = 2.16. (d) TR = 2.88. (e) TR = 3.60. The shaded regions represent one standard deviation.

# Supporting Information for: Modeling state-transition dynamics in resting-state brain signals by the hidden Markov and Gaussian mixture models

Takahiro Ezaki[1,2], Yu Himeno[3], Takamitsu Watanabe[4,5], and Naoki Masuda[6,7]

[1]Research Center for Advanced Science and Technology, The University of Tokyo, 4-6-1 Komaba, Meguro-ku, Tokyo 153-8904, Japan
[2]PRESTO, JST, 4-1-8 Honcho, Kawaguchi, Saitama 332-0012, Japan
[3]Department of Aeronautics and Astronautics, The University of Tokyo, 7-3-1 Hongo, Bunkyo-ku, Tokyo 113-8656, Japan
[4]Laboratory for Cognition Circuit Dynamics, RIKEN Centre for Brain Science, Saitama 351-0198, Japan
[5]International Research Center for Neurointelligence, The University of Tokyo 7-3-1 Hongo Bunkyo-ku, Tokyo 113-0033, Japan
[6]Department of Mathematics, State University of New York at Buffalo, Buffalo, NY 14260-2900, USA
[7]Computational and Data-Enabled Science and Engineering Program, State University of New York at Buffalo, Buffalo, NY 14260-5030, USA

July 12, 2021

**Supplementary Figures**

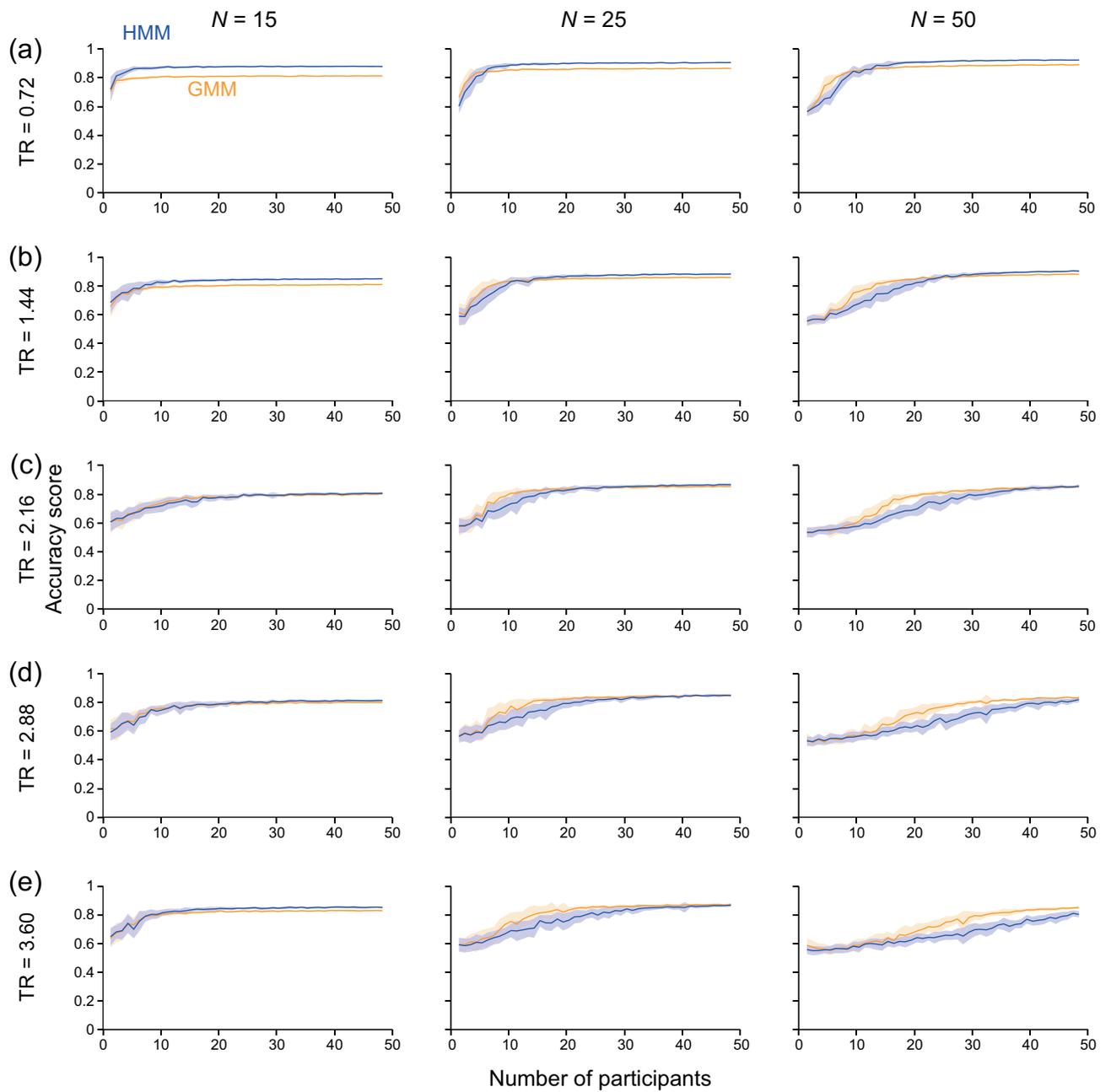

Supplementary Figure 1. Accuracy of estimating the hidden state for the GMM-based synthetic time series for $T = 5$ min. (a) TR = 0.72. (b) TR = 1.44. (c) TR = 2.16. (d) TR = 2.88. (e) TR = 3.60. The shaded regions represent one standard deviation.

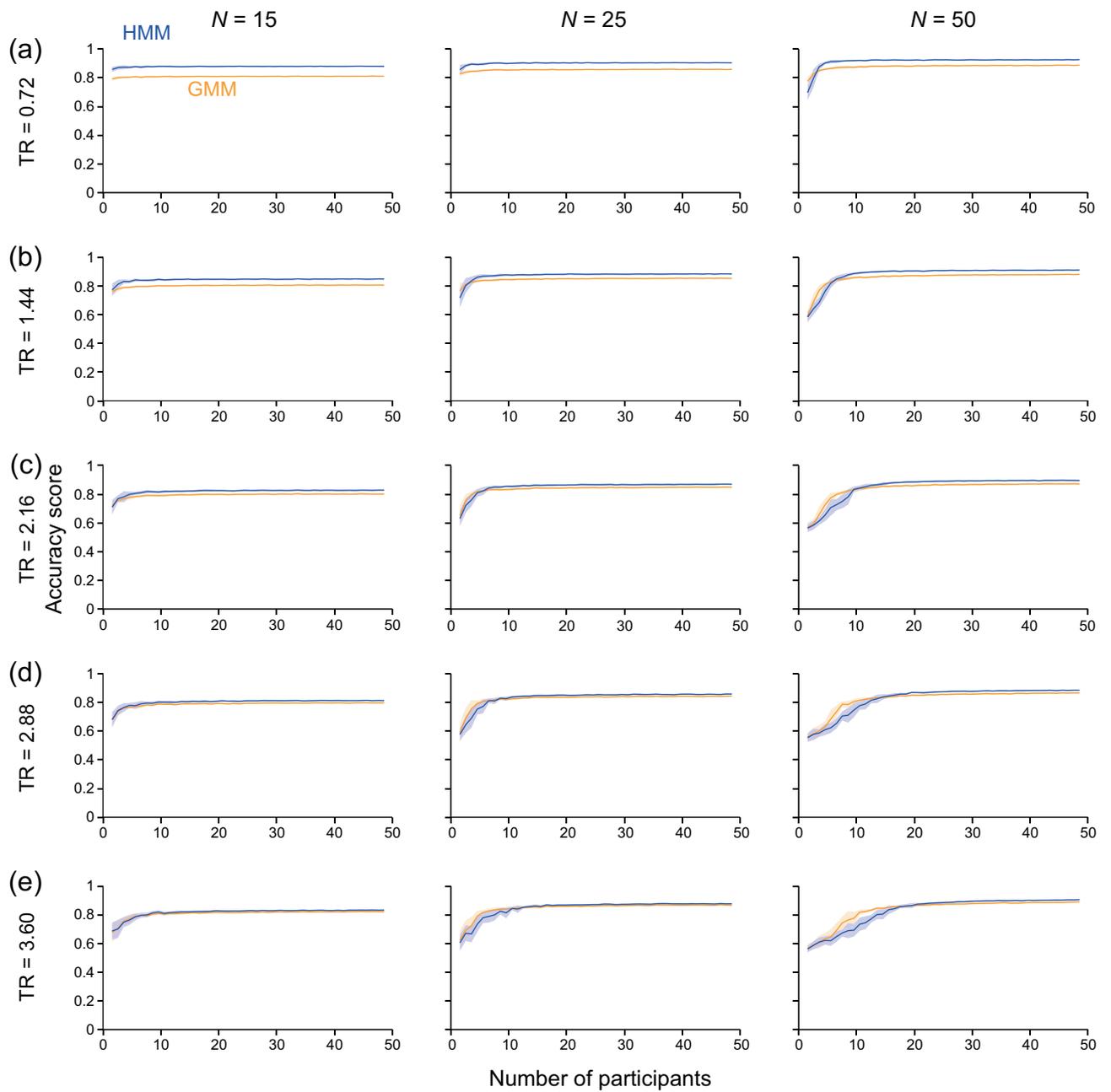

Supplementary Figure 2. Accuracy of estimating the hidden state for the GMM-based synthetic time series for $T = 14.4$ min. (a) TR = 0.72. (b) TR = 1.44. (c) TR = 2.16. (d) TR = 2.88. (e) TR=3.60. The shaded regions represent one standard deviation.

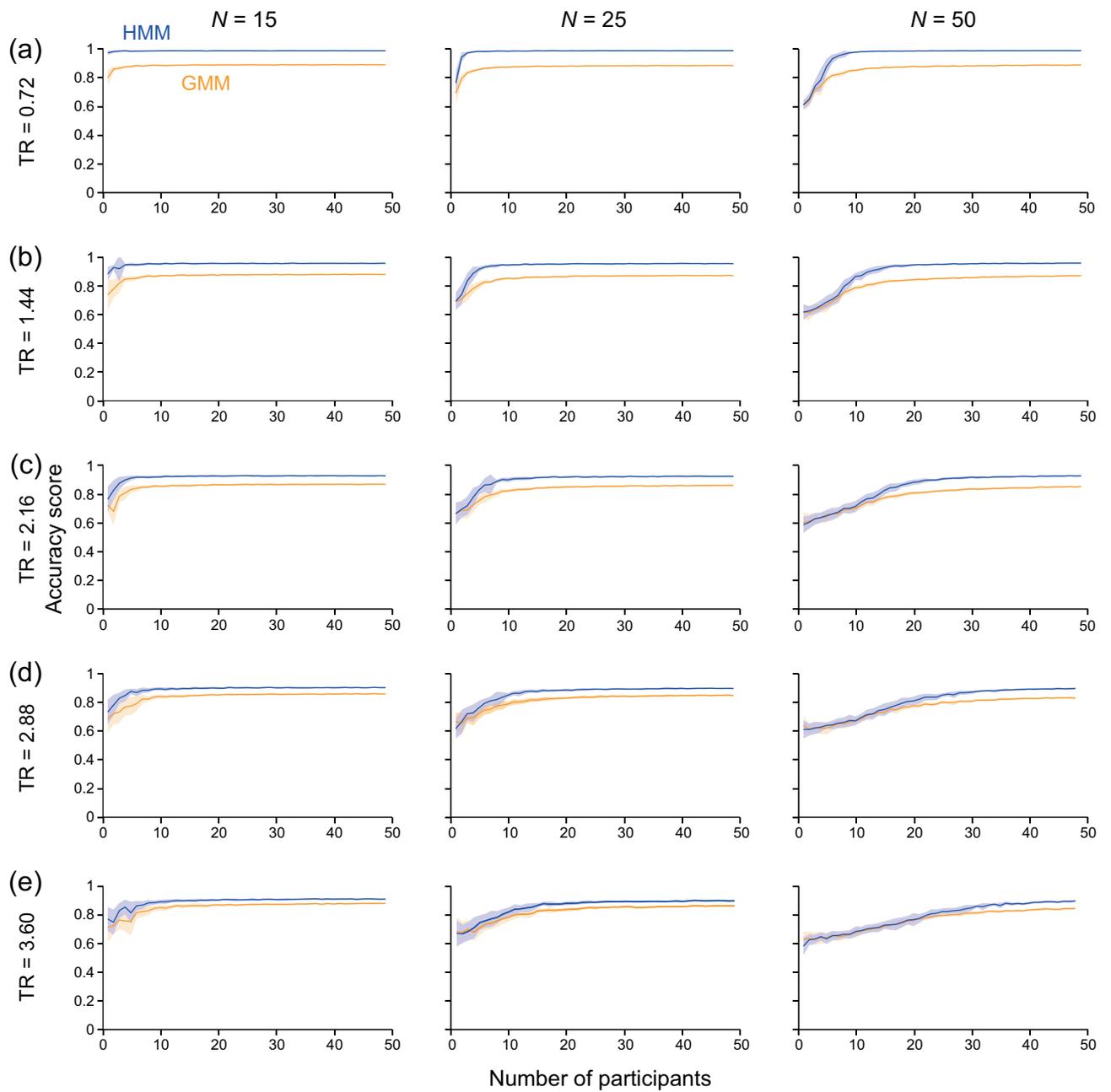

Supplementary Figure 3. Accuracy of estimating the hidden state for the HMM-based synthetic time series for $T = 5$ min. (a) TR = 0.72. (b) TR = 1.44. (c) TR = 2.16. (d) TR = 2.88. (e) TR=3.60. The shaded regions represent one standard deviation.

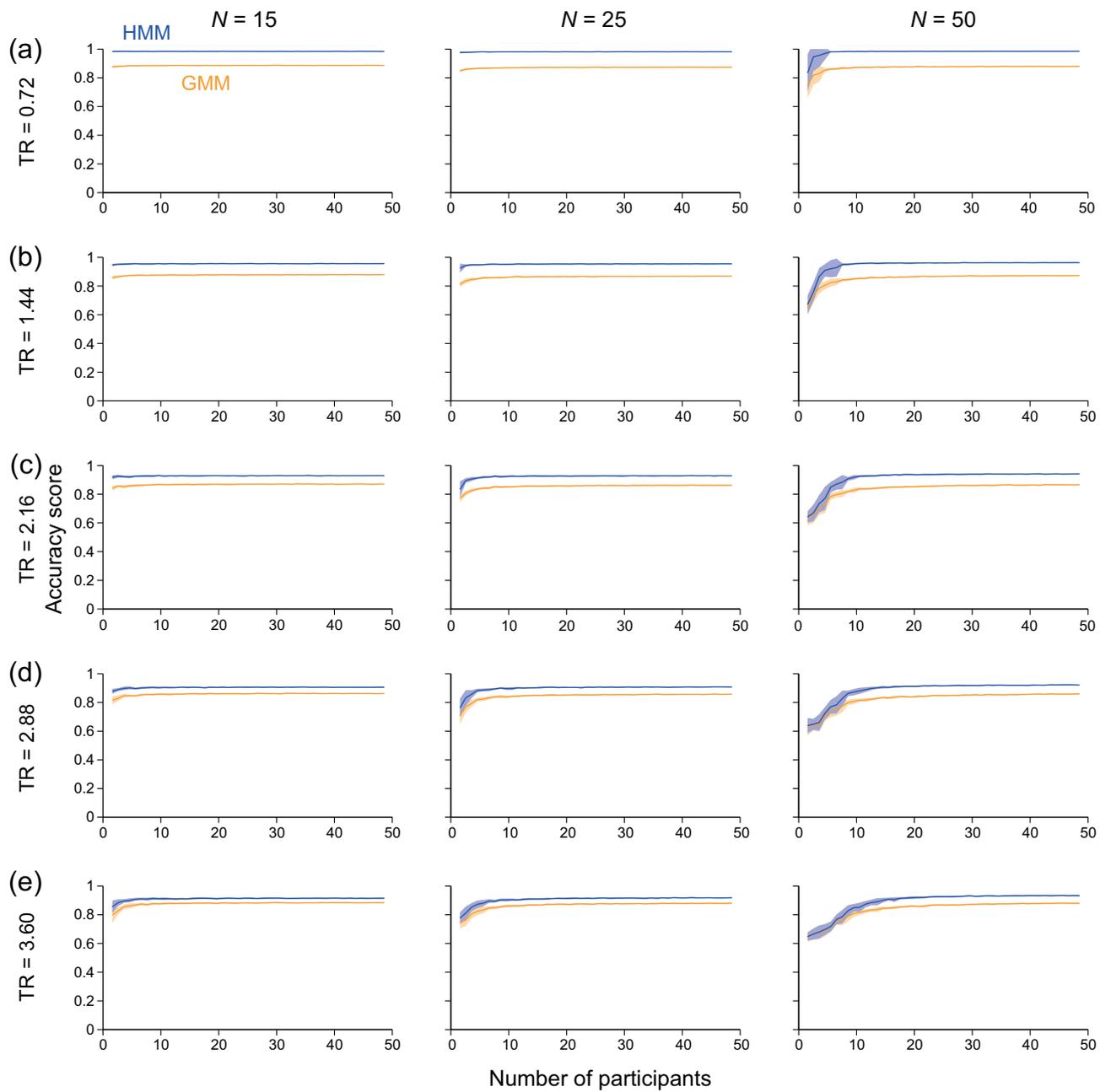

Supplementary Figure 4. Accuracy of estimating the hidden state for the HMM-based synthetic time series for $T = 14.4$ min. (a) TR = 0.72. (b) TR = 1.44. (c) TR = 2.16. (d) TR = 2.88. (e) TR=3.60. The shaded regions represent one standard deviation.